\documentclass{article}
\usepackage{spconf,amsmath,graphicx}
\usepackage{amssymb}
\usepackage{multirow}
\usepackage{hyperref}
\title{Regularized Contrastive Pre-training for Few-shot \\Bioacoustic Sound Detection}
\twoauthors
    {Ilyass Moummad, Nicolas Farrugia}
    {IMT Atlantique, Lab-STICC, \\UMR CNRS 6285, Brest, France,}
    {Romain Serizel}
    {University of Lorraine, CNRS, Inria, \\Loria, 54000, Nancy, France,}
\begin{document}
%\ninept
%
\maketitle
\begin{abstract}
Bioacoustic sound event detection allows for better understanding of animal behavior and for better monitoring biodiversity using audio. Deep learning systems can help achieve this goal. However, it is difficult to acquire sufficient annotated data to train these systems from scratch. To address this limitation, the Detection and Classification of Acoustic Scenes and Events (DCASE) community has recasted the problem within the framework of few-shot learning and organize an annual challenge for learning to detect animal sounds from only five annotated examples. In our study, we introduce a regularization to supervised contrastive loss, to learn non redundant features that exhibit effective transferability to few-shot tasks involving the detection of animal sounds not encountered during the training phase. Our method achieves a high F-score of 61.52\%$\pm$0.48 when no feature adaptation is applied, and an F-score of 68.19\%$\pm$0.75 when we further adapt the learned features for each new target task.  This work aims to lower the entry bar to few-shot bioacoustic sound event detection by proposing a simple and yet effective framework for this task, and by providing open-source code.\footnote{\url{https://github.com/ilyassmoummad/RCL_FS_BSED}}
\end{abstract}
\let\thefootnote\relax\footnotetext{This work is co-funded by the AI@IMT program of the ANR (French National Research Agency) and the company OSO-AI.}
%\vspace{-0.4cm}
\section{Introduction}
%\vspace{-0.2cm}
\label{sec:intro}
Bioacoustics delve into the study of sound production, emission, reception, and processing in living organisms. This diverse domain encompasses a wide range of research, from understanding the vocalizations of marine life to deciphering the intricate communication patterns of various animal species. Given the abundance and complexity of acoustic data in bioacoustics, the application of deep learning techniques has emerged as a powerful approach to extract meaningful insights from this soundscape~\cite{stowell2022}. 
% Deep learning, a subfield of machine learning, has demonstrated success across various domains, especially in computer vision, natural language processing and audio signal processing. Its ability to automatically learn intricate patterns and representations from raw data makes it an attractive candidate for bioacoustics research. Leveraging deep learning methodologies in bioacoustics promises to unlock a wealth of knowledge previously hidden within the vast acoustic repertoire of the natural world~\cite{stowell2022}.
% The integration of deep learning in bioacoustic research offers several advantages. Firstly, it allows for efficient and automated analysis of large-scale acoustic datasets, enabling researchers to process vast amounts of audio recordings more accurately and expeditiously. Secondly, deep learning algorithms can discern subtle patterns and features in bioacoustic signals that might not be immediately evident to human analysts. This capability significantly enhances the resolution and precision of acoustic pattern recognition, leading to more profound discoveries and insights into the behavior and communication of organisms~\cite{nolasco2023}.

Despite the considerable successes of deep learning in bioacoustics, there exists a significant challenge that hinders its widespread applicability – the scarcity of labeled data~\cite{stowell2022}. Annotating acoustic data is a laborious and time-consuming task that requires expertise in the understanding of the species. Consequently, available labeled bioacoustic datasets are often limited in size, impeding the full potential of data-hungry deep learning models. It is in this context that "few-shot bioacoustics" emerges as a promising area of research~\cite{nolasco2023}. 

Few-shot learning (FSL) is a subfield of machine learning that aims to train models using only a limited number of labeled examples. In the context of bioacoustics, this translates to developing robust and effective deep learning models that can generalize from a small number of annotated recordings, alleviating the data scarcity challenge. By harnessing few-shot learning techniques, researchers can circumvent the need for massive labeled datasets, making bioacoustic analyses more feasible for lesser-known species or habitats where extensive annotated data is lacking.

While FSL offers a compelling solution to mitigate the data scarcity challenges in bioacoustics, the effectiveness of these models heavily relies on the quality of the learned representations. In this context, representation learning plays a pivotal role in shaping the success of FSL-based approaches. A good starting initialization is crucial for FSL, and this is where representation learning techniques, like contrastive learning (CL)~\cite{simclr}, come into play.

CL is a learning paradigm designed to learn a metric space where similar samples are pulled together while dissimilar samples are pushed apart. CL has been widely used in the litterature and has shown promising results in audio representation learning~\cite{uclser}. However, CL can have the dimensional collapse phenomenon, where embedding vectors collapse along certain dimensions, thus only spanning a lower-dimensional subspace~\cite{collapse}. 

We propose a system that learns good intialization for FSL using supervised contrastive pre-training. To remedy the dimensional collapse of CL, we constrain the learned features to be diverse and non-redundant, using a regularization from information theory literature~\cite{mcrr}. Our goal is to learn features that are discriminative, ideally features that can cover a space of the largest possible dimension~\cite{mcrr}. %parler un peu de regularization% 

We apply the above pre-training strategy to train a general feature extractor for bioacoustic few-shot sound event detection (BSED). At inference, the feature extractor is either used directly for fast inference or fine-tuned for each binary validation task, specific to each audio file, for to the presence or absence of the event of interest, utilizing a prototypical loss. To make predictions, we slide a window over the audio file and compute an euclidean distance between the representations of each query window and the two prototypes (computed by averaging the representation of the annotated segments of presence/absence of the event of interest). We demonstrate the effectiveness of our approach on the diverse bioacoustic validation datasets of the DCASE challenge, showcasing its ability to achieve remarkable performance on the few-shot setting. 

This work builds upon our previous work~\cite{moummad}, where we pre-trained a feature extractor using CL and then trained a linear classifier on the available shots. While this system was the second best one in the challenge, the training of linear classifier using cross-entropy resulted in instability in some validation runs due to the large imbalance between the segments for the presence and absence of an event. Here, we replace the cross-entropy classification with a robust metric approach that is more stable and that optionally adapts the features to the task at hand. Additionally, we further enhance the pre-training stage by regularizing the learned representations. %Our main contribution in this work is regularizing supervised contrastive learning for learning transferable features, as well as proposing a simple and effective method for few-shot bioacoustic sound detection. %Our findings contribute to the advancement of few-shot bioacoustic research underscoring the efficacy of robust transfer learning.%
%\vspace{-0.4cm}
\section{Related Work}
\label{sec:related}
%\vspace{-0.2cm}
% We highlight in this section various works in the literature of bioacoustic sound event detection and classification, few-shot in this context, as well as representation learning. 

The DCASE community propose a benchmark for BSED that consists in detecting animal vocalizations in audio recordings given only five annotated examples~\cite{nolasco2023}. Liu et al.~\cite{surrey} use prototypical networks on the concatenation of per-channel energy normalization and delta mel-frequency cepstral coefficients, and trained on extra animal data from AudioSet~\cite{audioset} to increase generalization. Tang et al.~\cite{tang2022} use a frame-level approach using semi-supervised learning to exploit unlabeled query data. Our previous work~\cite{moummad} shows the strong performance of supervised contrastive pre-training followed by cross-entropy linear classification. Yan et al.~\cite{yan2023} improve over their previous work~\cite{tang2022} by adding target speaker voice activity detection to form a multi-task frame-level system, and by adding a transformer encoder in their model architecture.

MetaAudio~\cite{metaaudio} is a few-shot audio classification benchmark with diverse audio types (including bioacoustics). Our work doesn't address classification and reserves it for future research. BirdNet~\cite{birdnet}, a deep learning system trained on diverse data sources to identify 984 bird species, and Google Perch\footnote{https://tfhub.dev/google/bird-vocalization-classifier/4}, another model trained on an extensive bird corpus, have shown superior transferability for few-shot bioacoustic classification tasks when compared to models trained on generic audio datasets such as AudioSet~\cite{audioset}, as demonstrated by Ghani et al.~\cite{ghani2023}.

%This work focuses on the type of source dataset used for transfer learning and not the representation learning method.
%The previous works focus on the data used for training, we instead focus on the learning method.
The litterature of representation learning has shown great transfer performance thanks to CL~\cite{simclr, scl, uclser}. %However, CL can cause dimensional collapse~\cite{collapse}.
Regularized methods constrain the embeddings to have non-redundant information by measuring the cross-correlation between the representations of two views~\cite{bt}, decorrelating the feature variables from each other~\cite{vicreg}, or by maximizing the total coding rate of the features~\cite{emp, mcrr}. The combination of contrastive and regularized methods has not been yet explored. We investigate them in the context of transfer learning for few-shot bioacoustic sound event detection.

\begin{table*}[!h]
\centering
\caption{Performance on the validation datasets.}
%%\vspace{0.1cm}
\label{Tab:perf}
\scalebox{0.8}{%
\begin{tabular}{l|c|c|c|ccc|ccc|ccc}
\hline
\multirow{2}{*}{System} & \multirow{2}{*}{Precision} & \multirow{2}{*}{Recall} & \multirow{2}{*}{F1-score} & \multicolumn{3}{c|}{HB}                                         & \multicolumn{3}{c|}{ME}                                         & \multicolumn{3}{c}{PB}                                          \\ \cline{5-13} 
                        &                            &                         &                           & \multicolumn{1}{c|}{Pr}    & \multicolumn{1}{c|}{Re}    & F1    & \multicolumn{1}{c|}{Pr}    & \multicolumn{1}{c|}{Re}    & F1    & \multicolumn{1}{c|}{Pr}    & \multicolumn{1}{c|}{Re}    & F1    \\ \hline
%\multicolumn{13}{c}{\textit{Baseline}}  \\ \hline
\multicolumn{13}{c}{No extra data} \\ \hline
Template Matching           & 2.42                      & 18.32                   & 4.28                     & \multicolumn{1}{c|}{-} & \multicolumn{1}{c|}{-} & - & \multicolumn{1}{c|}{-} & \multicolumn{1}{c|}{-} & - & \multicolumn{1}{c|}{-} & \multicolumn{1}{c|}{-} & - \\
ProtoNets           & 36.34                      & 24.96                   & 29.59                     & \multicolumn{1}{c|}{-} & \multicolumn{1}{c|}{-} & - & \multicolumn{1}{c|}{-} & \multicolumn{1}{c|}{-} & - & \multicolumn{1}{c|}{-} & \multicolumn{1}{c|}{-} & - \\ %\hline
Moummad et al.~\cite{moummad}          & \textbf{73.93}                      & 55.59                   & 63.46                     & \multicolumn{1}{c|}{\textbf{82.95}} & \multicolumn{1}{c|}{82.32} & \textbf{82.63} & \multicolumn{1}{c|}{\textbf{67.69}} & \multicolumn{1}{c|}{84.61} & \textbf{75.21} & \multicolumn{1}{c|}{\textbf{72.72}} & \multicolumn{1}{c|}{33.33} & 45.71 \\
%\hline
%\multicolumn{13}{c}{Ours (using only the annotated segments)} \\
%\hline
\multirow{2}{*}{No fine-tuning (Ours)} & 60.99 & 62.08 & 61.52 & \multicolumn{1}{c|}{75.81} & \multicolumn{1}{c|}{78.00} & 76.89 & \multicolumn{1}{c|}{54.94} & \multicolumn{1}{c|}{92.95} & 69.04 & \multicolumn{1}{c|}{56.36} & \multicolumn{1}{c|}{40.48} & 47.11 \\
& $\pm$0.58 & $\pm$1.21 & $\pm$0.48 & \multicolumn{1}{c|}{$\pm$1.16} & \multicolumn{1}{c|}{$\pm$1.21} & $\pm$1.10 & \multicolumn{1}{c|}{$\pm$2.36} & \multicolumn{1}{c|}{$\pm$0.96} & $\pm$2.03 & \multicolumn{1}{c|}{$\pm$2.16} & \multicolumn{1}{c|}{$\pm$1.97} & $\pm$1.97
\\
\multirow{2}{*}{Fine-tuning (Ours)} & 65.00 & \textbf{71.75} & \textbf{68.19} & \multicolumn{1}{c|}{74.63} & \multicolumn{1}{c|}{\textbf{85.11}} & 79.52 & \multicolumn{1}{c|}{58.12} & \multicolumn{1}{c|}{\textbf{95.73}} & 72.30 & \multicolumn{1}{c|}{64.44} & \multicolumn{1}{c|}{\textbf{51.01}} & \textbf{56.93} \\
& $\pm$1.19 & $\pm$1.22 & $\pm$0.75 & \multicolumn{1}{c|}{$\pm$1.21} & \multicolumn{1}{c|}{$\pm$2.33} & $\pm$1.58 & \multicolumn{1}{c|}{$\pm$2.48} & \multicolumn{1}{c|}{$\pm$1.86} & $\pm$1.97 & \multicolumn{1}{c|}{$\pm$1.97} & \multicolumn{1}{c|}{$\pm$1.40} & $\pm$1.24 \\
\hline
% \multicolumn{13}{c}{Extra data (use of AudioSet Strong in training) / Semi-supervised (use of training data for each audio file fine-tuning)} \\
\multicolumn{13}{c}{Extra data} \\
\hline
Liu et al.~\cite{surrey}         & \textbf{76.56}                      & 49.54                   & 60.16                     & \multicolumn{1}{c|}{97.95} & \multicolumn{1}{c|}{79.46} & \textbf{87.74} & \multicolumn{1}{c|}{86.27} & \multicolumn{1}{c|}{84.62} & 85.44 & \multicolumn{1}{c|}{57.52} & \multicolumn{1}{c|}{27.66} & 37.36 \\
Tang et al. (SL)~\cite{tang2022}          & -                      & -                   & 66.6                     & \multicolumn{1}{c|}{-} & \multicolumn{1}{c|}{-} & 85.8 & \multicolumn{1}{c|}{-} & \multicolumn{1}{c|}{-} & 79.2 & \multicolumn{1}{c|}{-} & \multicolumn{1}{c|}{-} & 48.1 \\
Yan et al. (FL)~\cite{tang2022,yan2023}          & 73.0                      & 67.6                   & 70.2                     & \multicolumn{1}{c|}{-} & \multicolumn{1}{c|}{-} & 77.0 & \multicolumn{1}{c|}{-} & \multicolumn{1}{c|}{-} & 90.0 & \multicolumn{1}{c|}{-} & \multicolumn{1}{c|}{-} & 53.7 \\
Yan et al. (MTFL)~\cite{yan2023}          & 76.2                      & \textbf{75.3}                   & \textbf{75.7}                     & \multicolumn{1}{c|}{-} & \multicolumn{1}{c|}{-} & 86.7 & \multicolumn{1}{c|}{-} & \multicolumn{1}{c|}{-} & \textbf{90.2} & \multicolumn{1}{c|}{-} & \multicolumn{1}{c|}{-} & \textbf{58.9} \\
\hline 
\multicolumn{13}{c}{\scriptsize *We highlight in bold the best score for each metric.}
\end{tabular}
}
\end{table*}
%\vspace{-0.4cm}
\section{Method}
\label{sec:method}
\begin{figure}[!h]
\begin{minipage}[b]{1.0\linewidth}
  \centering
  \centerline{\includegraphics[width=1.\textwidth]{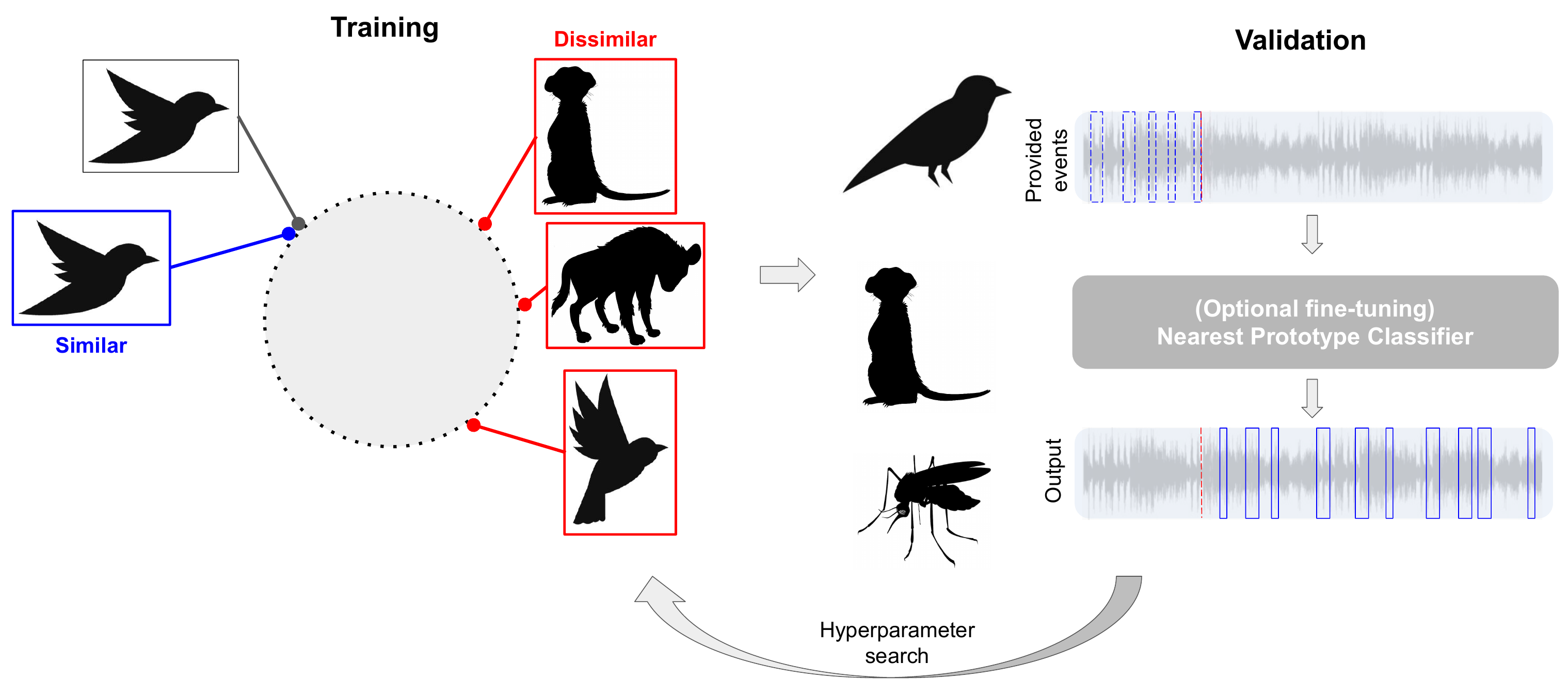}}
%  %\vspace{2.0cm}
\end{minipage}
\caption{Overview of our approach: Supervised contrastive pre-training, optionally fine-tuning the features, followed by nearest prototypical classifier.}
\label{fig:pip}
\end{figure}

%\vspace{-0.2cm}
In this section we describe the methodology employed in our study (Fig.~\ref{fig:pip}). We train a feature extractor on a general, labeled training set using supervised contrastive learning (SCL) combined with a coding rate regularization that constrains the embeddings to be non-redundant. The resulting trained model is transferred to the validation sets and optionally fine-tuned on the available shots using a prototypical loss. The predictions are made by computing the distances to the positive and negative prototypes, for the presence and absence of sound events of interest, respectively.

\vspace{-0.6cm}
\subsection{Supervised Contrastive Learning}
\label{sec:scl}
%\vspace{-0.2cm}
SCL consists in learning an embedding space in which the samples with the same class labels are close to each other, and the samples with different class labels are far from each other. Formally, a composition of an encoder $f$ and a shallow neural network $h$ called a projector (usually a MLP with one hidden layer) are trained to minimize the distances between representations of samples of the same class while maximizing the distances between representations of samples belonging to different class. After convergence, $h$ is discarded, and the encoder $f$ is used for transfer learning on downstream tasks. SCL loss is calculated as follows:
\begin{equation}
    \label{scl}
    \mathcal{L}^{SCL} = \sum_{i\in I}\frac{-1}{|P(i)|}\sum_{p\in P(i)}\log{\frac{\text{exp}\left(\boldsymbol{z}_i\boldsymbol{\cdot}\boldsymbol{z}_p/\tau\right)}{\sum\limits_{n\in N(i)}\text{exp}\left(\boldsymbol{z}_i\boldsymbol{\cdot}\boldsymbol{z}_n/\tau\right)}}
\end{equation}
where $i\in I$ is the index of an augmented sample within a training batch, containing two views of each original sample. These views are constructed by applying a data augmentation function $A$ twice to the original samples. $\boldsymbol{z}_{i}=h(f(A(\boldsymbol{x}_{i})))\in\mathbb{R}^{D_P}$ where ${D_P}$ is the projector's dimension. ${P(i)={\{p\in I:{{y}}_p={{y}}_i}\}}$ is the set of indices of all positives in the two-views batch distinct from $i$ sharing similar label with $i$. $|P(i)|$ is its cardinality, $N(i)=I\setminus \{i\}$, the $\boldsymbol{\cdot}$ symbol denotes the dot product, and $\tau\in\mathbb{R}^{+*}$ is a scalar temperature parameter.
%\vspace{-0.2cm}
\subsection{Regularization : Total Coding Rate}
\label{sec:tcr}
%\vspace{-0.2cm}
In Information Theory, the coding rate is the proportion of bits that carry non-redundant information. Let $Z = [z_1, ..., z_b]$ be a batch of $b$ features of dimension $d$. The total coding rate (TCR)~\cite{emp} $\mathcal{R}$ of $Z$ is defined as follows: 
\begin{equation}
    \label{tcr}
    \mathcal{R}(Z) = \frac{1}{2}\log \text{det}\left(I + \frac{d}{b \epsilon^2}ZZ^T\right)
\end{equation}
where $\epsilon>0$ is a chosen precision. The training loss is:
\begin{equation}
    \label{trloss}
    \mathcal{L}^{Train} = {L}^{SCL} - \lambda {R}(Z)
\end{equation}
where $\lambda > 0$ is a hyperparameter coefficient for the regularization term.
We want the coding rate of $Z$ to be as large as possible. The TCR regularization can be seen as a soft-constrained regularization of covariance term in VICReg~\cite{vicreg}, where the covariance regularization is achieved by maximizing TCR~\cite{emp}.
%\vspace{-0.4cm}
\subsection{Fine-tuning}
\label{sec:proto}
%\vspace{-0.2cm}
Using the same annotations as section~(\ref{sec:scl}), we define the fine-tuning loss as:
\begin{equation}
    \label{proto}
    \mathcal{L}^{Finetune} = -\log\frac{\text{exp}\left(\boldsymbol{z}_i\boldsymbol{\cdot}\boldsymbol{z}_c\right)}{\sum\limits_{c\prime\neq c}\text{exp}\left(\boldsymbol{z}_i\boldsymbol{\cdot}\boldsymbol{z}_{c\prime}\right)}
\end{equation}
% référencer papiers domain shift feature adaptation (mieux motiver protonets)
This loss is similar to the ProtoNets loss~\cite{protonets}, which produces a distribution over classes for a query point based on a softmax over distances to the prototypes in the embedding space. However, we do not do meta-testing using episodes as in ProtoNets, we instead do regular batch training by fine-tuning the model using the augmented batch similarly to the supervised contrastive pre-training stage. We slightly modify the ProtoNets loss by removing the distance to the corresponding prototype from the summation in the denominator. Our intuition is drawn from the work of DCL~\cite{dcl}, which enhanced performance by removing the positive comparison from the denominator of the normalized temperature-scaled cross-entropy loss (NT-Xent) originally used in SimCLR~\cite{simclr}(Eq.\ref{simclr}). 
\begin{equation}
    \label{simclr}
    \mathcal{L}^{SimCLR} = -\log\frac{\text{exp}\left(\boldsymbol{z}_i\boldsymbol{\cdot}\boldsymbol{z}_{i\prime}\right)}{\sum\limits_{j\neq i,i\prime}\text{exp}\left(\boldsymbol{z}_i\boldsymbol{\cdot}\boldsymbol{z}_{j}\right)}
\end{equation}
We observe that in the NT-Xent loss (Eq. \ref{simclr}), when substituting the second element of each similarity term with the corresponding prototype, we obtain the $\mathcal{L}^{Finetune}$ loss. %For validation tasks we optionally fine-tune the pre-trained model using $\mathcal{L}^{Finetune}$.
%\vspace{-0.4cm}
\subsection{Nearest Prototype Classifier}
\label{sec:proto}
%\vspace{-0.2cm}
To make predictions, for each audio file, we compute the Euclidean distances between the queries and the prototypes to assign the labels of presence/absence of the event of interest. For robustness, each segment (both query and prototype) is augmented to create multiple views.
%using random resized time crop of ratio sampled uniformly between 90\% and 100\% of the total duration, and a power gain of coefficient sampled uniformly between 0.9 and 1. This data augmentation procedure is lighter than the one performed during pre-training (section~\ref{sec:da}) as our goal here is not to learn invariance to transformation but to create multiple views with small noise.%
The representations of these views are averaged to one representation vector, in addition, the positive and negative segments are also averaged to have one positive and one negative prototypes. Using the annotations from subsection(~\ref{sec:tcr}), let ${Z_i}$ be the subset of ${Z}$ with class label $i$, we then define the prototype ${\Bar{\mathcal{Z}_i}}$ for each class label $i$ as: 
\begin{equation}
    \label{prototypes}
    \forall i:\Bar{\mathcal{Z}_i} = \frac{1}{|Z_i|}\sum\limits_{z\in {Z_i}} z
\end{equation}
Let $q$ be a query, we predict its label $i_q$ as:
\begin{equation}
    \label{prediction}
    i_q= \arg\min_i \|q-\Bar{\mathcal{Z}_i}\|_2
\end{equation}
The onsets and offsets decision of the event of interest is made based on the precise moment when the label for the next query transitions from a negative class to a positive class and from a positive class to a negative class, respectively.
%\vspace{-0.4cm}
\section{Experiments}
We experiment on the BSED datasets from DCASE and refer the reader to the work of Nolasco et al.~\cite{nolasco2023} for more details about these datasets.
\subsection{Model Backbone}
%\vspace{-0.2cm}
Our architecture is the same as the one used in our previous work~\cite{moummad}. We use a ResNet consisting of three blocks (64$\rightarrow$128$\rightarrow$256), each comprising three convolutional layers. We employ max pooling operations after each block of a kernel of size 2x2 for the first and second blocks, and of size 1x2 for the third block.
%\vspace{-0.4cm}
\subsection{Training and validation procedure}
%\vspace{-0.2cm}
We train our model from scratch on the training set using SCL framework with a temperature of 0.06, regularized with TCR with a square precision of 0.05 and a regularization coefficient of 0.001. We use SGD optimizer with a batch size of 128, a learning rate of 0.01 with a cosine decay schedule, momentum of 0.9, and a weight decay of 0.0001 for 100 epochs. We use the data augmentation policy in table~\ref{Tab:da}.

%\vspace{-0.4cm}
\begin{table}[!h]
\caption{Training data augmentations. \textbf{SM:} Spectrogram Mixing, \textbf{FS:} Frequency Shift, \textbf{RRTC:} Random Resized Time Crop, \textbf{PG:} Power Gain, \textbf{AWGN:} Additive White Gaussian Noise.}
\centering
\scalebox{0.9}{%
\begin{tabular}{l|c|c|c|c|c}
%\hline
\label{Tab:da}
Augs   & SM & FS & RRTC & PG & AWGN\\
\hline
Params & factor & bands & ratio & factor & std\\
Values & $\beta(5,2)$ & {[}0-10{]} & {[}0.6,1.0{]} & {[}0.75-1{]} & {[}0-0.1{]} 
\end{tabular}
}
%\hline
\end{table}

During the validation phase, we optionally fine-tune the whole model using $L^{Finetune}$ for adapting the features for each audio recording using a learning rate of 0.01 for 40 epochs. For this purpose, we used random resized time crop (RRTC) of ratio sampled uniformly between 90\% and 100\% of the total duration, and power gain (PG) of coefficient sampled uniformly between 0.9 and 1. This data augmentation procedure is lighter than the one performed during pre-training (\ref{Tab:da}), and is also used to create multiple views for each query window during inference. In all our experiments, we train the backbone with three different seeds, and for each backbone, we conduct three evaluations, resulting in a total of 9 runs per experiment.
%\vspace{-0.4cm}
\section{Results}
Table~\ref{Tab:perf} shows our results, the baseline and the first two ranking teams of the 2022 and 2023 DCASE challenge editions. Our method outpeforms that of Liu et al.\cite{surrey} (both with and without fine-tuning). We also improve upon our previous work~\cite{moummad} with fine-tuning. While Yan et al.\cite{tang2022} and Tang et al.\cite{yan2023} achieve better results with their semi-supervised frame-level (FL) approach, we outperform their segment-level (SL) approach. For a fair comparison, we divide Table~\ref{Tab:perf} into methods that utilize extra data (such as AudioSet Strong~\cite{surrey} or the reuse of training data for the adaptation of features on each audio recording~\cite{tang2022, yan2023}) and those that do not. We note that our approach utilizes only the available shots during inference, making it practical for real-time applications or settings with limited resources.
% We also note that unlike these two works, our approach does not use unlabeled query data, which make it practical for real-time applications.
In Table~\ref{Tab:abla}, we study pre-training strategies without fine-tuning, showing the superiority of regularized SCL (+TCR) compared to vanilla SCL, SimCLR and Cross-Entropy. In Table~\ref{Tab:ablaft}, we analyze fine-tuning methods : SCL, original Prototypical Loss, and $\mathcal{L}^{Finetune}$, confirming insights about removing the positive comparison from the denominator of the prototypical loss.% (Table~\ref{Tab:abla}).

\begin{table}[!h]
\centering
\caption{Ablation of the pre-training method w/o fine-tuning.}
\label{Tab:abla}
\centering
\scalebox{0.9}{%
\begin{tabular}{l|c|c|c}
\hline
Method & Precision & Recall & F1-score
\\ \hline
Cross-Entropy & 34.59$\pm$1.21 & 62.35$\pm$0.93 & 44.49$\pm$1.22 \\
SimCLR & 54.75$\pm$0.77 & 61.16$\pm$1.62 & 57.75$\pm$0.41 \\
%SCL & 55.38$\pm$2.65 & 62.93$\pm$0.75 & 58.89$\pm$1.74 \\
SCL & 56.80$\pm$2.98 & \textbf{62.77$\pm$0.77} & 59.59$\pm$1.75 \\
SCL+TCR & \textbf{60.99$\pm$0.58} & 62.08$\pm$1.21 & \textbf{61.52$\pm$0.48} \\ \hline
%SCL+TCR (fine-tune) & - & - & - \\ \hline
\end{tabular}
}
\end{table}

\vspace{-0.6cm}

\begin{table}[!h]
\caption{Ablation study on the fine-tuning method.}
\centering
\label{Tab:ablaft}
\scalebox{0.9}{%
\begin{tabular}{l|c|c|c}
\hline
Method & Precision & Recall & F1-score
\\ \hline
SCL & 62.75$\pm$1.34 & 70.92$\pm$0.72 & 66.58$\pm$1.05 \\
Original Proto & 55.62$\pm$2.68 & \textbf{72.13$\pm$0.67} & 62.77$\pm$1.86 \\
$\mathcal{L}^{Finetune}$ & \textbf{65.00$\pm$1.19} & 71.75$\pm$1.22 & \textbf{68.19$\pm$0.75} \\ \hline
%SCL+TCR (fine-tune) & - & - & - \\ \hline
\end{tabular}
}
\end{table}
\vspace{-0.2cm}

\section{Conclusion}
%\vspace{-0.2cm}
In this work, we have presented a simple yet effective approach for bioacoustic few-shot sound event detection. Our approach involves pre-training a feature extractor using supervised contrastive learning with a regularization that enforces learning non-redundant features. The feature space learned by our approach allows for computing directly distances to the prototypes for making prediction. We also propose to further enhance the performance by fine-tuning the features for each audio file at the cost of longer inference. 
For our future work, we want to generalize our approach to bioacoustic sound event classification and explore robust feature adaptation techniques for when fewer shots are available (one-shot). We will also explore the frame-level approach, as well as a proposal-based approach for detecting variable length temporal regions of interest, that have not been previously investigated in this task.

\vfill\pagebreak

\bibliographystyle{IEEEbib}
\bibliography{refs}

\end{document}